\documentclass[prb,twocolumn, superscriptaddress,preprintnumbers,amsmath,amssymb]{revtex4}

\usepackage[usenames,dvipsnames]{xcolor}

\usepackage[normalem]{ulem}
 \newcommand{\stkout}[1]{\ifmmode\text{\sout{\ensuremath{#1}}}\else\sout{#1}\fi}

\usepackage{float}
\usepackage{graphicx}
\usepackage{dcolumn}
\usepackage{bm}
\usepackage{multirow}
\usepackage{amsmath,amssymb}

\begin{document}

\title{U 5$f$ crystal-field ground state of UO$_2$ probed by \\ directional dichroism in nonresonant inelastic x-ray scattering}

\author{M.~Sundermann}
   \affiliation{Institute of Physics II, University of Cologne, Z{\"u}lpicher Stra{\ss}e 77, D-50937 Cologne, Germany}
   \affiliation{Max-Planck-Institute for Chemical Physics of Solids - N{\"o}thnizer Stra{\ss}e 40, 01187 Dresden, Germany}
    
\author{G.~\surname{van~der~Laan}}
\email[Corresponding author: ]{gerrit.vanderlaan@diamond.ac.uk}
       \affiliation{Magnetic Spectroscopy Group, Diamond Light Source, Didcot OX11 0DE, United Kingdom}

\author{A.~Severing}
    \affiliation{Institute of Physics II, University of Cologne, Z{\"u}lpicher Stra{\ss}e 77, D-50937 Cologne, Germany}
    \affiliation{Max-Planck-Institute for Chemical Physics of Solids - N{\"o}thnizer Stra{\ss}e 40, 01187 Dresden, Germany}

\author{L.~Simonelli}
   \altaffiliation{Present address:\ ALBA Synchrotron Light Source, E-08290 Cerdanyola del Vall\`{e}s, Barcelona, Spain}
	 \affiliation{European Synchrotron Radiation Facility (ESRF), B.P. 220, 38043 Grenoble C\'edex, France}
\author{G.~H.~Lander}
   \affiliation{European Commission, Joint Research Centre (JRC), Directorate for Nuclear Safety and Security, Postfach 2340, D-76125 Karlsruhe, Germany}	
\author{M.~W.~Haverkort}
    \affiliation{Institute for Theoretical Physics, Heidelberg University, Philosophenweg 19, 69120 Heidelberg, Germany}
\author{R.~Caciuffo}
    \affiliation{European Commission, Joint Research Centre (JRC), Directorate for Nuclear Safety and Security, Postfach 2340, D-76125 Karlsruhe, Germany}

\date{\today}

\begin{abstract}
Nonresonant inelastic x-ray scattering (NIXS) has been performed on single crystals of UO$_2$ to study the direction dependence of higher-order-multipole scattering from the uranium  $O_{4,5}$ edges (90--110 eV). By comparing the experimental results with theoretical calculations the symmetry of the ground state is confirmed directly as the crystal-field (CF) $\Gamma_5$ triplet state within the $J$ = 4 manifold. The results also demonstrate that the directional dichroism of the NIXS spectra is sensitive to the CF strength and establish NIXS as a tool for probing CF interactions quantitatively.
\end{abstract}


\maketitle


Whilst in low energy x-ray spectroscopy  the electric-dipole transitions  prevail, the advent of high-energy synchrotron radiation has provided the opportunity to explore also higher-order multipole transitions. These transitions are governed by different selection rules that are reaching other final states, offering complementary spectroscopic perspectives. 
This is especially useful because the different multipole transitions are often well-separated in energy from each other due to the electrostatic interactions in the final state.
Furthermore, the bulk sensitivity of hard x-rays offers a clear advantage over soft x-rays, and  allows spectroscopy under extreme conditions as well as the study of samples, such as actinides, that must be encapsulated for safety reasons.

One such high-energy technique is nonresonant inelastic x-ray scattering (NIXS), in which a photon-in ($\nu_i$), photon-out ($\nu_f$) process  promotes a core electron  to an unoccupied valence state. 
This is given by the transition $\ell^n + \nu_i \to \underline{c} \ell ^{n+1} + \nu_f$, where $\ell^n$ represents a valence shell $\ell$ with $n$ electrons, and $ \underline{c}$ denotes a hole in the core state.
 Multipole moments $k$ for the $c \to \ell$ transition are allowed by the triangle condition $| \ell -c| \le k \le \ell +c$ and parity rule $ \ell + c + k$ = even. Thus for $d \to f$ transitions, $k$ = 1 (dipole), $k$ = 3 (octupole), and $k = 5$ (triakontadipole) transitions are allowed. 
 The  relative contributions  of the multipole moments  depend on the value of the momentum transfer $\mathbf{q}$.
A high intensity of the $k$ = 3 and 5 transitions compared to the $k$ = 1 is obtained by using a large $|\mathbf{q} |$  ($\sim$ 10\,\AA$^{-1}$), as achieved with hard x rays, typically $\sim$10\,keV, at large scattering angles.    
NIXS has been  well described theoretically and  successfully compared to experiments assuming spherical symmetry.  \cite{Soininen2005,Larson2007,Haverkort2007,Gordon2008,Gordon2009,Bradley2010,Caciuffo2010,Bradley2011,Laan2012prl,Laan2012prb,Willers2012,Sundermann2016,Sundermann2018} 
In this Letter we investigate the angular dependence.
 
NIXS has no intermediate state so that the interpretation is as straightforward as for x-ray absorption spectroscopy (XAS). \cite{Laan2014} The momentum transfer {\bf{q}} in NIXS takes the place of the light polarization $\varepsilon$ in XAS. Importantly, XAS is almost exclusively governed by dipole transitions, which cannot distinguish between spherical and cubic symmetry, since  the transferred  angular momentum $k$ = 1 branches to a single irreducible representation.  However,
in NIXS the transitions with angular momentum $k$ = 3 and $k$ = 5 split   in cubic symmetry into three and four different irreducible representations, respectively. Measurements with $\hat{\mathbf{q}} = \mathbf{q}/ | \mathbf{q} |$ along different crystal directions therefore can give a nonzero difference signal (directional dichroism) that provides information on the non-sphericity of the electronic ground state.
NIXS measurements done by Gordon {\it et al.} \cite{Gordon2009}  at the Mn $M_{2,3}$ edge ($3p \to 3d$) on a cubic single crystal of MnO already revealed   different spectra  for $\hat{\mathbf{q}}$ along  [111] and [100] directions.

Here we demonstrate the power of the $\bf{q}$-direction dependence of NIXS in a single crystal of uranium dioxide. UO$_2$ has been studied extensively for more than 50 years. We know that this material has two $5f$ electrons with a $\Gamma_5$ triplet ground state. This information has primarily come from neutron scattering --- the crystal-field (CF) splitting was determined in 1989 \cite{Amoretti1989} and the excitation spectrum was reported conclusively in 2011 \cite{Caciuffo2011}, after pioneering work 50 years ago. \cite{Cowley1968} Extensive theory on the interactions in UO$_2$ \cite{Santini2009}   and more recently   studies using self-consistent DFT+$U$ calculations and a model Hamiltonian  \cite{Zhou2011,Zhou2012} have also been reported. Given this profound knowledge of the ground state of UO$_2$, we have performed  NIXS experiments to test that the new technique gives the correct results for UO$_2$, establishing the importance of the direction dependence of the NIXS spectra.  Comparing experiments with a series of calculations shows unambiguously that the $\Gamma_5$  triplet is indeed the ground state. In addition, by extending such calculations to cover different CF  strengths, we show that the technique is also sensitive to the magnitude of the crystal field.

\begin{figure*}[t]
   \centering
   \includegraphics[width=1.99\columnwidth]{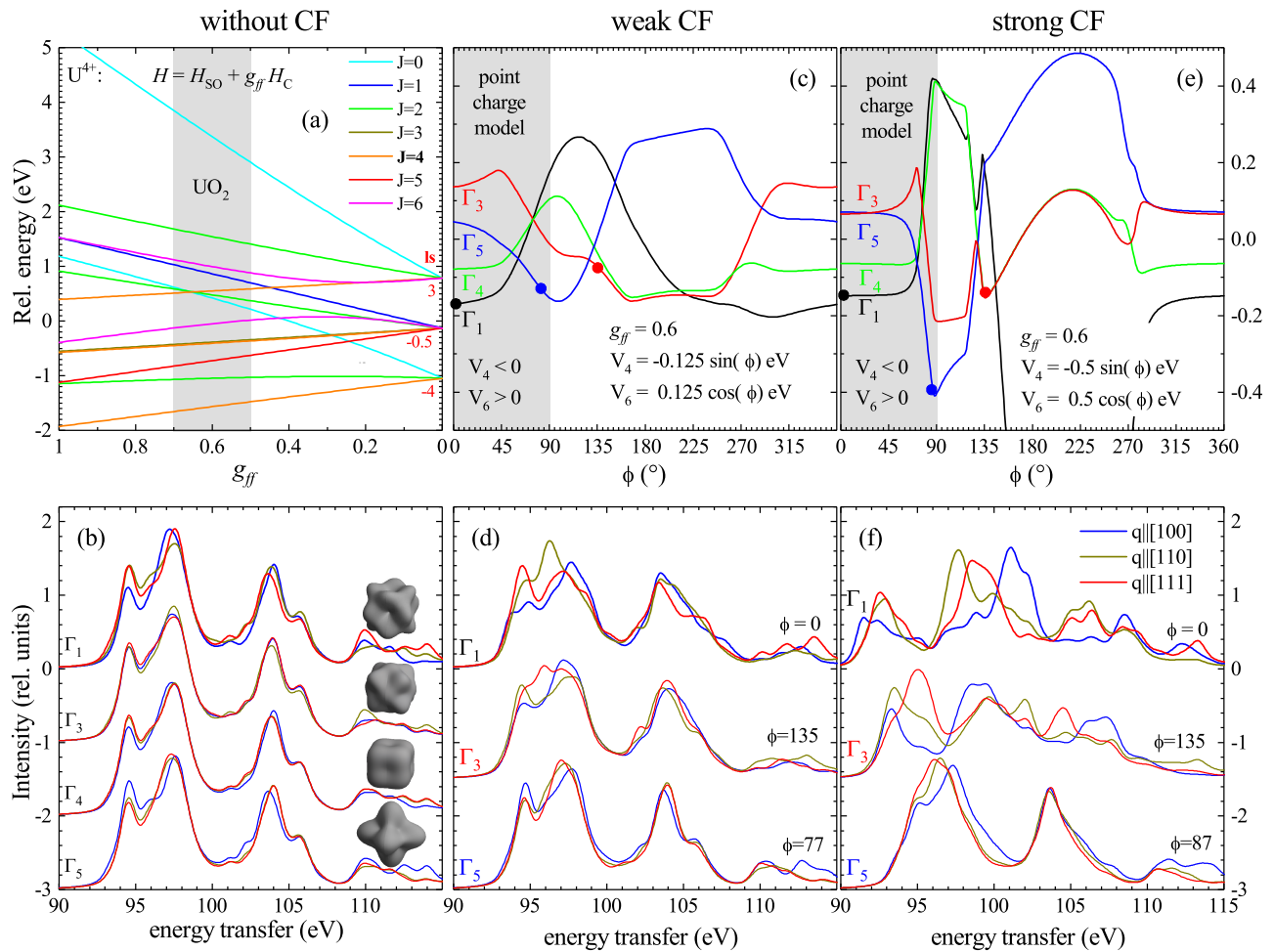}
   \caption{(a) Energy level diagram of the $J$ levels for the configuration U$^{4+}$ $5f^2$  without CF as  a function of the Slater reduction factor $g_{ff}$. 
   The grey-shaded region marks the applicable range for UO$_2$. (b) Calculated NIXS spectra of the four cubic eigenstates as constructed by symmetry for pure $J$=4 [see Eq.~(1)] for ${\hat{\bf{q}}}$$\|$[001] (blue), ${\hat{\bf{q}}}$$\|$[011] (dark yellow), and ${\hat{\bf{q}}}$$\|$[111] (red). The insets show the respective charge densities for two electrons. (c) and (e) energy-level diagrams of CF states as a function of the ratio of the CF parameters $V_4$ and $V_6$ expressed in terms of $\phi$, for a weak and strong CF scenario (see text). The colored dots mark selected $\Gamma_1$, $\Gamma_3$, and $\Gamma_5$    ground states used for the direction dependent NIXS calculations in panel (d) and (f). }
   \label{fig1}
\end{figure*}


NIXS measurements with  momentum transfer  of $|{\bf{q}}|$\,=\,9.1\,\AA$^{-1}$ were done on beamline ID20 at the European Synchrotron Radiation Facility (ESRF) on UO$_2$ crystals with (001) and (111) surfaces. 
Details of the  experimental setup  and crystals can be found in the Appendix.

\begin{figure}
   \centering
   \includegraphics[width=0.9\columnwidth]{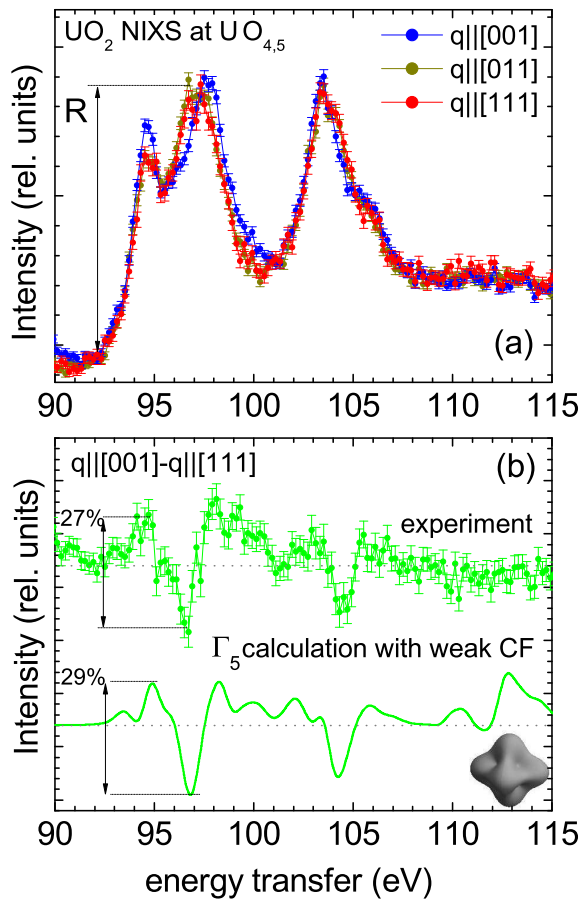}
   \caption{(a) NIXS data of UO$_2$ measured at 300\,K (b) Measured difference spectra (directional dichroism) of the two directions ${\hat{\bf{q}}}$$\|$$[001]  -  {\hat{\bf{q}}}$$\|$$[111]$ and corresponding calculation for the weak CF case with   $\Gamma_5$ ground state (green lines), see text.}
   \label{fig2}
\end{figure}

The (e$^{-1}$) penetration length for x rays of 10 keV into UO$_2$ is $\sim$5\,$\mu$m, so this probe  is sensitive to the top $\sim$2.5\,$\mu$m. This is far greater than any surface effects extend, so NIXS may truly be considered as a bulk probe. Data have been collected at room temperature by scanning the incident energy $E_{i}$ = $E_{f}$ + $\hbar\omega_i$ at fixed final energy, covering the  energy-transfer interval corresponding to the uranium $O_{4,5}$ ($5d \to 5f$)  absorption edges. The obtained results were consistent with earlier measured angular integrated spectra of UO$_2$. \cite{Caciuffo2010,Bradley2010}

Simulations were performed using the full-multiplet code {\texttt{Quanty}}\,\cite{Haverkort2016} that includes Coulomb and spin-orbit interactions. The calculations are based on an ionic approach for the U$^{4+}$\,5$f^2$ configuration. In intermediate coupling, the total momentum $J$ is a good quantum number and the ground state is  $J$\,=\,4  [Fig.\,\ref{fig1}(a)]. The atomic parameters were calculated using  Cowan's atomic multiplet code\,\cite{Cowan1981} and the Slater integrals for Coulomb interaction were reduced to account for intra-atomic relaxation effects.\cite{Moore2009} Figure \ref{fig1}(a) shows the energy-level diagram for U$^{4+}$\,5$f^2$ as a function of the Slater reduction factor $g_{ff}$.
This reduction factor was adjusted such that the energy distribution of the calculated isotropic and experimental pseudo-isotropic spectra matched. The latter was constructed from the weighted sum of the measured directions; the calculated spectrum is the sum of the diagonal elements of the scattering tensor (see Appendix). Here no CF has been taken into account. For the $5f$--$5f$ and $5d$--$5f$ Coulomb interactions we found reduction factors  $g_{ff}$  and $g_{fc}$ that were both equal to 0.6, and the 5$f$ spin-orbit coupling was not reduced. The reduction factors change the overall shape of the isotropic spectra but have no direct influence on the $\bf{q}$-direction dependence of spectra. For the simulations we used a slightly larger value of $|{\bf{q}}|$ than given by the experimental scattering triangle because the radial part of the wave functions is based on the atomic values. A variation of $|{\bf{q}}|$ changes slightly the ratio of the multipole contributions. \cite{Caciuffo2010,Sundermann2016} Finally, a Gaussian and  Lorentzian broadening of 0.65  and 0.6\,eV FWHM, respectively, accounted for instrumental resolution and life-time effects.


If the Coulomb interaction is much larger than the spin-orbit interaction ($LS$-coupling limit) the ground state is $^3H_4$, but a finite spin-orbit interaction give intermixing with the 
$^1G_4$ and $^3F_4$ states.
 In the  Stevens approximation, \cite{Stevens1952} which ignores intermixing with $J$\,$\neq$\,4 levels, the ground state is a mixture of different $J$\,=\,4 levels (see Table\,I in the Appendix). The CF acts only on $J$ and can be written written for the cubic point-group $O_h$  in the $|J_z\rangle$ basis set.  Defining the $\hat{J}_x$ operator for the phase relation positive, the ninefold $J$ = 4 level splits up into
\begin{align}
\begin{array}{lrlcrlcrl}
 |\Gamma_1 \rangle =
 & 0.456 & |-$$4\rangle & + &  0.456 & |+$$4\rangle & + & 0.764 & |0\rangle , \\ \\
\multirow{2}{*}{$ |\Gamma_3 \rangle = \left\{ \vphantom{\begin{array}{l}~\\~\end{array}} \right. $}
 & 0.541 & |-$$4\rangle & + & 0.541 & |+$$4\rangle & - & 0.644 & |0\rangle , \\
 & 0.707 & |-$$2\rangle & + & 0.707 & |+$$2\rangle , \\ \\
\multirow{2}{*}{$ |\Gamma_4 \rangle = \left\{ \vphantom{\begin{array}{l}~\\~\end{array}} \right. $}
 & 0.707 & |-$$4\rangle & - & 0.707 & |+$$4\rangle , \\
 & 0.935 & |\mp1\rangle & + & 0.355 & |\pm3\rangle , \\ \\
\multirow{2}{*}{$ |\Gamma_5 \rangle = \left\{ \vphantom{\begin{array}{l}~\\~\end{array}} \right. $}
 & 0.707 & |-$$2\rangle & - & 0.707 & |+$$2\rangle , \\
 & 0.355 & |\mp$$1\rangle & - & 0.935 & |\pm$$3\rangle  . 
 \end{array}
\end{align}

We   calculated the NIXS spectra of these eigenstates for different ${\hat{\bf{q}}}$ directions [Fig.\,\ref{fig1}(b)]. The respective two-electron charge densities are displayed as insets.  Note that these are the \textsl{`calculations without CF'} since the states were constructed as given above, i.e.\ in the absence of finite CF. The spectra with ${\hat{\bf{q}}}$$\|$[001] (blue) and ${\hat{\bf{q}}}$$\|$[111] (red) in Fig.\,\ref{fig1}(b) show the strongest direction dependence, especially at   $\sim$95, 97, and 105\,eV. Particularly, the $\Gamma_5$ and $\Gamma_1$ show an opposite direction dependence at these energies. Hence it should be possible to identify a $\Gamma_5$ state, which is the CF ground state expected from previous neutron inelastic scattering study.\cite{Amoretti1989}


Experimental results are shown in Fig.\,\ref{fig2}(a).
While it is tempting to assign the main peak splitting in the  spectra as $O_5$ ($5d_{5/2} \to 5f)$ and $O_4$ ($5d_{3/2}  \to  5f$),   caution should be exercised.
In XAS and electron-energy-loss spectroscopy (EELS) the dipole transitions ($k$ = 1)  result in a broad peak around 110 eV accompanied by a small prepeak at $\sim$105 eV  with an energy splitting that is mainly governed by the $5d$--$5f$ exchange interaction. \cite{Moore2009}
In NIXS, on the other hand, the observed energy splitting between the (95--97)  and 105 eV peaks is primarily due to the $5d$  core spin-orbit splitting.  \cite{Laan2012prb}
The peak at 95 eV arises mainly from $k$ = 5 transitions, whereas those at 97 and 105 eV arise from both $k$ = 3 and $k$ = 5 transitions. According to the spin-orbit sum rule  \cite{Laan2012prl} a  change in  angular momentum quantum number $J$   changes the   intensity ratio of the  spin-orbit-split peaks. Here instead we will be looking for   subtle differences in the angular dependence.

For the actinide $O_{4,5}$, as well as the rare-earth $N_{4,5}$,  edges the dipole transitions are strongly broadened due to super-Coster-Kronig decay to   continuum states. \cite{Terry2002,Moore2009}    
However, compared to the dipole transitions the higher-order multipole transitions, which excite to final states with larger spin and orbital momenta, are shifted towards  lower energy due to the strong $5f$--$5d$ exchange interaction. \cite{Laan2012prb} As a result,   higher-order multipole transitions have   narrower line width with a broadening primarily determined by the core-hole lifetime.

Figure\,\ref{fig2}(a) shows the experimental NIXS spectra for the same momentum transfer and directions as in Fig.\,\ref{fig1}(b), i.e.\ for ${\hat{\bf{q}}}$$\|$[001] (blue), ${\hat{\bf{q}}}$$\|$[011] (dark yellow), and ${\hat{\bf{q}}}$$\|$[111] (red). Comparing the data and simulations in  Figs.\,\ref{fig2}(a) and \ref{fig1}(b)   shows that only the $\Gamma_5$ calculation matches the direction dependence of the experimental data.

So far we ignored the intermixing with $J$\,$\neq$\,4 levels, although $J$ mixing is expected in actinide compounds. The question is to what extent this mixing affects the interpretation of the NIXS data. We therefore calculated NIXS spectra for different CF scenarios. Figure\,\ref{fig1}(c) and \ref{fig1}(e) show the energy-level diagrams for a weak and strong CF potential as a function of the ratio of the CF parameters $V_4$ and $V_6$. 
Their ratio is expressed in terms of $\phi$ with $V_4 = -A \sin\phi$ and $V_6 = A \cos\phi$ with $A = 0.125$ and  0.5 for the weak and strong CF case, respectively. 
In the grey-shaded range ($\phi$$<$$90^\circ$), $V_4$$<$0 and $V_6$$>$0 
according to a point-charge model. \cite{Lea1962} The wild $\phi$ dependence of the energy levels in the strong CF case is due to the large $LSJ$  intermixing, which in the weak CF case is much more behaved. It turns out that $\Gamma_4$ can never be the ground state, and neither can $\Gamma_3$  within the point-charge model, in agreement with Lea, Leask, and Wolf.\cite{Lea1962} For the remaining states the direction dependent NIXS spectra have been calculated.

The NIXS spectra for selected CF parameters that produce a $\Gamma_1$ (black dots), $\Gamma_3$ (red dots), and $\Gamma_5$ (blue dots)   ground state  are shown in Fig.\,\ref{fig1}(d) and  \ref{fig1}(f). The scheme of Amoretti \textsl{et al.}\,\cite{Amoretti1989} is realized in the weak CF scenario for $\phi$\,=\,77$^{\circ}$ and that of   Rahman\,\&\,Runciman\,\cite{Rahman1966} in the strong CF case for $\phi$\,=\,87$^{\circ}$. For the NIXS calculation the reduction factors and line widths are kept unchanged because modifying $g_{ff}$ and $g_{fc}$ does not improve agreement between calculated and measured pseudo-isotropic spectra (see Appendix).  Especially the spectra of the $\Gamma_1$ and $\Gamma_3$ ground states change substantially with increasing $LSJ$ intermixing, the $\Gamma_5$ lesser so. The mixing factors of the respective ground states are listed in Table I of the Appendix.

Comparison of the calculated CF ground state and data shows that the $\Gamma_1$ and $\Gamma_3$ still have to be excluded as ground states, both in the weak and strong CF scenarios. The $\Gamma_5$ of the strong CF case still shows some resemblance of the measured spectra, e.g., the high energy tail of the peak at $\sim$98\,eV still shows stronger scattering for ${\hat{\bf{q}}}$$\|$[001] than for the two other directions (blue over red and dark yellow). However, the direction dependence at 95\,eV is much better reproduced for the weak CF with a $\Gamma_5$ ground state. The latter actually describes the data very well. Figure\,\ref{fig2}(b) shows the excellent agreement of the measured and calculated direction dependence by comparing the difference plots of the [001] and [111] directions. Even the size of the observed dichroism agrees very well with the calculated one, as shown by the relative values of 27\% and 29\%. These numbers refer to the difference of the two directions at energies indicated in Fig.\,\ref{fig2}(b), relative to the peak height $R$ [as defined in Fig.\,\ref{fig2}(a)].

We should emphasise that the theoretical dichroism appearing above 110 eV is difficult to observe experimentally because all states at higher energy transfers, i.e.\ in the energy range of the dipole transition ($\hbar$$\omega$\,$>$\,108 eV),\,\cite{Caciuffo2010,Sundermann2016} appear strongly broadened due to interaction with continuum states. This increase in lifetime broadening
 is not captured in the calculations.\cite{Sundermann2018}



The exceptional agreement observed in Fig.\,\ref{fig2}(b) between experiment and theory shows that the symmetry is unambiguously that of the $\Gamma_5$ triplet in the weak CF scenario, i.e.\ within the $J$\,=\,4 manifold. Other CF symmetries as well as the strong CF scenario can be unambiguously excluded. 

Although CF transitions are observed in almost all rare-earth materials with neutron inelastic scattering,\cite{fulde85} this is not the case in actinide and some Ce heavy-fermion materials.\cite{holland94} For ionic materials, such as UO$_2$, CF-transitions are indeed observed,\cite{Amoretti1989} but for intermetallic systems the conduction electrons interact with the 5$f$ states to cause a severe broadening of the CF transitions. Also XAS and  EELS, in which  dipole transitions dominate, suffer from severe broadening for intermetallic uranium systems\,\cite{Moore2009,Jeffries2010,Wray2015} so that the excitonic effect of the higher multipole transitions in NIXS offers a great advantage. For UO$_2$, which is cubic, the dipole ($k$\,=\,1) spectrum cannot provide information on the anisotropy of the charge distribution, so that an examination of  higher-order multipole transitions is essential. {\it{Hence the NIXS technique represents an alternative method compared to neutron scattering and x-ray absorption spectroscopy to determine the symmetry of the ground state in such materials.}}
 Indeed, such a study  has already been published on {\it{tetragonal}} URu$_{2}$Si$_{2}$,\cite{Sundermann2016} and the ground-state symmetry was determined but without considering the effect of strong versus weak CF.
Neutron inelastic scattering, however, shows a number of broad CF transitions, and is thus unable to establish the ground state.\cite{santini00,park02}
\\

\begin{table*}[t]
\renewcommand{\arraystretch}{1.25}
	\begin{tabular*}{0.98\textwidth}{@{\extracolsep{\fill}}l|c||c|c||c|c||c|c}
		\hline
$LSJ$ states & Stevens& $\Gamma_5$ weak CF& $\Gamma_5$ strong CF & $\Gamma_3$ weak CF& $\Gamma_3$ strong CF &$\Gamma_1$ weak CF& $\Gamma_1$ strong CF \\ 
   	\hline
$^3H_4$&  0.860 &  0.864            &0.718 & 0.881  & 0.531 &  0.823   &   0.583   \\
$^3H_5$&   0    &  0.002            &0.010 &  0.011 & 0.105 &    0     &   0       \\
$^3H_6$&   0    &     0             &0.008 &  0.026 & 0.106 &   0.050  &   0.132   \\
$^3F_2$&   0    &    0              &0.047 &  0.013 & 0.101 &    0     &   0       \\
$^3F_3$&   0    &     0             &0.040 &   0    &  0    &     0    &   0       \\
$^3F_4$&  0.012 &  0.010            &0.022 &   0.044&  0.156&   0.116  &   0.270   \\
$^1G_4$&  0.128 &  0.122            &0.143 &  0.020 &  0    & 0.006    &   0       \\
$^1D_2$&   0    &    0              &0.002 &   0    &  0    &    0     &   0       \\
$^1I_6$&   0    &    0              &0.006 &   0.004&  0    &  0.003   &   0.003   \\
$^3P_0$&    0   &    0              &0     &  0     & 0     &   0.002  &   0.010   \\
$^3P_1$&    0   &    0              &0     &  0     &  0    &    0     &   0       \\
$^3P_2$&    0   &    0              &0.004 &  0.002 &  0    &    0     &   0       \\
$^1S_0$&    0   &    0              &0     & 0      &  0    &    0     &   0.002   \\

   	\hline
	\end{tabular*}
	\caption{Weights (squared contributions) of respective $LSJ$ states to the ground-state wave functions for the three different cases:\ Stevens, weak, and strong CF. In the case of the Stevens approximation the $J$ admixture is identical for all CF states. For the weak and strong case the ground-state wave functions are chosen as in Fig.\,1(d) and 1(f) of the main text.}
\end{table*}

\begin{figure}[t]
   \centering
   \includegraphics[width=0.9\columnwidth]{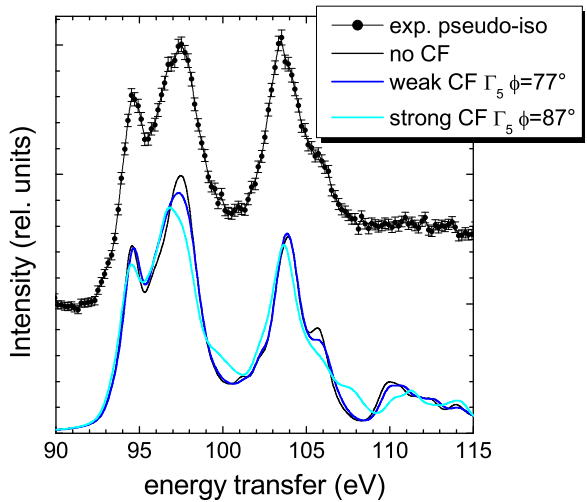}
   \caption{Experimental pseudo-isotropic spectrum of UO$_2$ as constructed from the single-crystalline data (black dots) and calculated NIXS spectra (lines), without considering a CF (black line), with a $\Gamma_5$ ground state in a weak (blue line), and strong CF (cyan).}
 \label{figS1}
\end{figure}

\section{Appendix}

  \subsection{Experimental details}

The NIXS experiment has been performed using the RIXS spectrometer on beamline ID20 at the European Synchrotron Radiation Facility in Grenoble, France.\cite{Moretti2018} The beam generated by three consecutive undulators was monochromatized  
using a channel-cut Si(311) coupled to
a Si(111) double-crystal momochromator, and horizontally focused by a Rh-coated mirror. A set of five spherically-bent Si(660) analyzer crystals with 1\,m bending radius, horizontal scattering geometry and vertical Rowland circles, provided an energy resolution of  $\sim$0.65\,eV at a final photon energy $E_{f}$ = 9.69002\,keV, and an intensity of 7\,$\times$\,10$^{13}$ photons/s for a 25\,$\mu$rad vertical divergence of the undulator radiation. The Bragg angle 
of the analyzers was fixed at 87.5$^\circ$. The analyzers were placed at scattering angles 2$\theta$\,=\,100, 110, 120, 130, and 140$^\circ$. The scattered intensity was recorded by a MAXIPIX fast readout, photon-counting position sensitive detector, achieving up to 1.4 kHz frame rate with 290 $\mu$s readout dead time, with a pixel size of 55 $\mu$m and a detection geometry of 256 $\times$ 256 pixels.

For the  measurements we used two UO$_{2}$ single crystals, originally cut and polished by Walt Ellis at Los Alamos National Laboratory \cite{taylor81}, with (001) and (111) surfaces, respectively, and fully described in Ref.\,\onlinecite{watson00}. The samples were aligned with the [001] and [111] surface normal parallel to the averaged momentum transfer that points towards 150$^\circ$ for elastic scattering. The [011] direction was realized by rotating the [001] crystal accordingly. For data analysis, only data collected in the highest analyzer at 140$^\circ$ were used because here the momentum transfer is largest ($|{\bf{q}}|$\,=\,9.1\,\AA$^{-1}$). The corresponding momentum transfer at elastic scattering points towards 160$^\circ$,  i.e.\  there is an offset of 10$^{\circ}$ between the respective crystallographic directions and ${\bf{q}}$ which has been taken into account in the data analysis.
\\
 
  \subsection{Calculation of the isotropic spectra}

The isotropic spectra have been calculated from the trace of the scattering tensor, which corresponds to an integration over  all  ${\hat{\bf{q}}}$ directions. The Gaussian linewidth
 is determined by the instrumental resolution, so that only the lifetime (Lorentzian linewidth), Slater reduction factors $g_{ff}$ and $g_{fc}$, and 5$f$ spin-orbit interaction are adjustable parameters. All calculations in 
Fig.\,\ref{figS1}
 are performed with the parameters as given in the main text.

The pseudo-isotropic spectrum is a linear combination of the three measured directions that yields an isotropic spectrum for $k$
\,=\,1 and  3, in good agreement with previously measured isotropic specra of UO$_2$. \cite{Caciuffo2010,Bradley2010}
The prefactors depend on the point group. For the present $O_h$ case: $I_{\mathrm{Iso}}\,=\,[40 \,I({\hat{\bf{q}}}\|[100])\,+\,32\, I({\hat{\bf{q}}}\|[110])\,+\,27 \,I({\hat{\bf{q}}}\|[111])]/99$ considering $k$\,=\,1 and 3. For $k$\,=\,5 transitions we did not measure enough directions to provide the true isotropic spectrum but the error is negligible when comparing calculations of the true and pseudo-isotropic spectra. 
\\

{\bf{Acknowledgment}}

We acknowledge beamtime on beamline ID20 at the ESRF in Grenoble under proposal HC-743.
M.S. and A.S. benefited from the financial support of the Deutsche Forschungsgemeinschaft (DFG) under grant SE-1441-5-1.

\bibliography{Biblio_G}
 
\end{document}